\documentclass[modern]{aastex631}
\usepackage[utf8]{inputenc}
\usepackage{amsmath}

\renewcommand{\paragraph}[1]{\medskip\par\noindent\textbf{#1}~---}
\sloppypar

\usepackage{graphicx}
\usepackage{xcolor}
\usepackage[framemethod=tikz]{mdframed}
\usetikzlibrary{shadows}
\definecolor{captiongray}{HTML}{555555}
\mdfsetup{%
  innertopmargin=2ex,
  innerbottommargin=1.8ex,
  linecolor=captiongray,
  linewidth=0.5pt,
  roundcorner=1pt,
  shadow=false,
}
\newlength{\figurewidth}
\shorttitle{combining spectra by forward modeling}
\shortauthors{hogg \& casey}
\newcommand{\documentname}{\textsl{Article}}
\newcommand{\sectionname}{Section}
\newcommand{\secref}[1]{\sectionname~\ref{#1}}
\newcommand{\name}{\textsl{Frizzle}}

\newcommand{\unit}[1]{\mathrm{#1}}

\newcommand{\kmps}{\unit{km\,s^{-1}}}

\begin{document}

\title{\large \name: Combining spectra or images by forward modeling}

\author[0000-0003-2866-9403]{David W. Hogg}
\affiliation{Center for Cosmology and Particle Physics, Department of Physics, New York University}
\affiliation{Max-Planck-Institut f\"ur Astronomie, Heidelberg}
\affiliation{Center for Computational Astrophysics, Flatiron Institute}

\author[0000-0003-0174-0564]{Andrew R. Casey}
\affiliation{School of Physics \& Astronomy, Monash University}
\affiliation{Centre of Excellence for Astrophysics in Three Dimensions (ASTRO-3D)}

\begin{abstract}\noindent
When there are many observations of an astronomical source---many images with different dithers, or many spectra taken at different barycentric velocities---it is standard practice to shift and stack the data, to (for example) make a high signal-to-noise average image or mean spectrum.
Bound-saturating measurements are made by manipulating a likelihood function, where the data are treated as fixed, and model parameters are modified to fit the data.
Traditional shifting and stacking of data can be converted into a model-fitting procedure, such that the data are not modified, and yet the output is the shift-adjusted mean.
The key component of this conversion is a spectral model that is completely flexible but also a continuous function of wavelength (or position in the case of imaging) that can represent any signal being measured by the device after any reasonable translation (or rotation or field distortion).
The benefits of a modeling approach are myriad:
The sacred data never are modified.
Noise maps, data gaps, and bad-data masks don't require interpolation.
The output can take the form of an image or spectrum evaluated on a pixel grid, as is traditional.
In addition to shifts, the model can account for line-spread or point-spread function variations, world-coordinate-system variations, and calibration or normalization variations.
The noise in the output becomes uncorrelated across neighboring pixels as the shifts deliver good coverage in some sense.
The only cost is a small increase in computational complexity over that of traditional methods.
We demonstrate the method with a small data example and we provide open-source sample code for re-use.
\end{abstract}


\section{Introduction}\label{sec:intro}

It would be wrong to say that the practice of astronomy is, basically, \emph{staring at the sky}.
That said, staring at the sky is a big part of what we do.
And the way we do it, usually, is this:
We take many individual images of our target with exposure times that are long enough to make read noise irrelevant (if possible) while at the same time short enough to avoid saturation (of what we care about).
We then shift and average (maybe median?) those images to get the highest possible signal-to-noise on our target.
This, fundamentally, is the technology underlying the \textsl{Hubble Deep Field} (\citealt{hdf}),
the \textsl{APOGEE} spectroscopic survey (\citealt{apogee}),
and the future Rubin Observatory \textsl{LSST} (\citealt{lsst}), among countless other projects.
This technology is encoded in tools like \texttt{Drizzle} (\citealt{drizzle}).

These traditional practices are reliable and fast.
However, they will---in some cases---sacrifice some information.
Minimum-variance unbiased estimators require care in their construction; they are generally maximum-likelihood estimators (\citealt{mvue}).
The deep, high signal-to-noise image or spectrum we seek is a statistical estimate of some kind of ``true'' or high signal-to-noise image or spectrum, given the noisy data of the individual exposures.
That is, a combined image is a kind of \emph{prediction}: It is the prediction of what \emph{would have been observed} with a much longer exposure with a much better device.
In measurement theory, the practice is to combine noisy data into a measurement by building and optimizing a likelihood.
Of course we don't always think of a combined (or mosaiced) image of the sky or of an average spectrum of a star as the result of applying an \emph{estimator}, but it is.
It is just a simultaneous estimate of a lot of different components---a lot of different pixels---all the pixels in the combined image or spectrum.

We are in a period in astrophysics in which exceedingly precise measurements are expected, in both imaging and spectroscopy projects.
In spectroscopy, in particular, high signal-to-noise spectral measurements are expected to obtain surface-abundance measurements with precisions a few percent, and radial-velocity measurements with precisions of up to one ten-thousandth of a \emph{pixel} (in terms of the Doppler shift).
Spectral combination methods that distort absorption (or emission) lines will translate into biases, errors, or increased variance in these measurements.
Spectral combination methods that correlate the noise in the output combined spectrum will translate into increased variance, and necessitate more sophisticated downstream measurement techniques.

One of the great things about the standard practice---the simple process of shifting and stacking the data---is that this procedure makes almost no assumptions about what the output, combined image or spectrum will look like.
It isn't really a ``model'' for the data; it is more like a ``summary'' of the data.
When we switch to a statistical-inference framework, we have to choose some form or basis for expressing the combined image.
In what follows we use a maximally parameterized model, in the sense that we choose an expression for the functions or function space that has as many parameters as the output image has pixels.
That means that the approach we take does not get its power from the choice of parameterized functions for the output:
Indeed it has just as much freedom as the shifting-and-stacking standard practice.
The forward-modeling we recommend gets its power from the fact that it generates the data directly and without manipulation, by means of a justifiable likelihood function.

A boundary condition of the work presented here is that the functional requirements of anything we provide must be the same as with standard practice:
That is, we consider here only procedures that start with individual, single-epoch observations as inputs, along with noise estimates and bad-pixel masks; and we consider only procedures that end with an output ``combined image'' or ``combined spectrum'' that is a set of intensity or flux values on a pixel grid, along with uncertainty estimates and metadata flags.
That is, the only difference between standard practice and the forward model that we advocate is purely in the nature of the estimator.
Unlike the shift-and-coadd approach, the estimator we will deliver will be insensitive to under- or poorly-sampled data, which is not uncommon when astronomical seeing accidentally gets excellent, or in spectrographs that have limited detector pixels (for example, \citealt{apogeehardware}).
As we will show with artificial data, the estimator we propose yields a faithful estimate of the \emph{true} spectrum, and usually with good noise properties.
Both of these results make our results good for subsequent precision measurements (e.g., radial velocities, chemical abundances).

The approach we propose differs from interpolation-and-averaging in a subtle way. 
Interpolation could be described as computing new values at arbitrary positions based on an existing set of known data points. 
Instead, here we propose to construct a continuous spectral model for the data, such that we can evaluate that model on either the input (raw-data) pixel grid or else on the output (combined spectrum) pixel grid.
The \emph{output product} is the same form as what one would expect from an interpolation-and-averaging procedure (values at a set of pixel positions), but the \emph{procedure} differs from interpolation. When interpolating the input data the basis functions are explicitly constructed from the input-data pixel grid, whereas here we construct a forward model that is maximally consistent with the data, and can be evaluated at any pixel locations.

The ideas here aren't novel; many projects have avoided data interpolation previously.
For one example, the \textsl{COMBO-17} survey (\citealt{combo17}) took many images of the sky in many bands, and did not co-add the data prior to making measurements, in part to avoid mixing different point-spread functions and pixel samplings.
For another, some projects have sought to co-add background-dominated images in a near optimal way, but presume that all observations are already registered and re-sampled to the same grid \citep{zackay}.
For another, the \textsl{DESI} survey ensures that all spectra are extracted on the same wavelength grid, to avoid any interpolation or resampling \citep{desipipeline}.
What we propose below is actually a sub-problem of the problem solved by the \textsl{wobble} method for determining stellar radial velocities (\citealt{wobble}); \textsl{wobble} is more general because it also includes a tellurics model and also tellurics and stellar variability models,
plus it learns (it isn't told) the epoch-to-epoch spectral shifts.
There is a heritage for what's presented here in the ``optimal extraction'' literature (for example, \citealt{oe}, \citealt{kelson}, \citealt{froe}, \citealt{sp}):
Optimal extraction describes a process of extracting a one-dimensional spectral signal from a two-dimensional spectrograph image when that extraction proceeds by forward modeling the two-dimensional data.
One amusing point, which is out of scope here, is that if the data that are being combined are extracted one-dimensional spectra, and those are, in turn, optimal extractions, then there is almost certainly an even better method that moves this spectral combination method directly into the two-dimensional data; that is, there ought to be a method that obviates one-dimensional spectral extraction entirely.
We hope for a future in which one-dimensional spectra are never extracted from two-dimensional images; in this future the two-dimensional data will be the sacred data, not the extractions.

In real applications, the spectra we extract are often of variable sources, or taken at different position angles on extended objects, such that we don't expect each epoch to be identical.
Similarly, the \textsl{LSST} data are being taken the way they are because the sky (and point-spread function) will vary.
Hence, in assuming that we only want the time-invariant mean of our data, we are considering here the very easiest problem in a range of possible problems.
However, we note that all standard shift-and-coadd methods make similar assumptions implicitly.
Part of our contribution in this \documentname{} is to make implicit assumptions explicit; we do this in \secref{sec:assumptions}.

Finally it is worthy of remark that, in some conceptual sense, there is no reason to \emph{ever} combine the individual spectroscopic exposures at all, for any reason, except possibly visualization and good vibes.
No information is gained by performing a combination.
Any line or feature that becomes visible under spectral combination must---at least in principle---be measurable in the uncombined individual spectra if those individual spectra are analyzed appropriately simultaneously.
That was the insight and motivation for the unusual \textsl{COMBO-17} (\citealt{combo17}) approach in which the individual exposures were never co-added or combined; it is also the underlying idea behind work we have done on the proper motions and parallaxes of very faint sources (\citealt{undetectable}).
Averaging the data doesn't ever create any new information.
It can be satisfying though; it meets astronomical community expectations; and it might provide insights or inspire discoveries.

\section{Concepts and assumptions}\label{sec:assumptions}

What we do here can apply to almost any astronomical signals, but we are going to specialize to multi-epoch astronomical spectra for specificity.
We are going to make many other assumptions too, all of which are discussed in this \sectionname.

\paragraph{Multi-epoch spectra with shifts}
We will assume that there are $N$ observations (epochs) of the same star (or maybe of different stars or quasars, say, that are assumed to be more-or-less identical, intrinsically, but let's think of a single star for now).
Each of these observations $i$ (with $1\leq i\leq n$) produces a one-dimensional observed (noisy) image $y_i$.
The $y_i$ can be thought of as $M_i$-vectors in what follows, where $M_i$ is the number of pixels in spectrum $i$.
The different observations are made at different relative velocities (star minus spectrograph) such that the images are shifted by a different shift $\Delta x_i$ relative to the detector or extracted wavelength grid.

\paragraph{Shift operators}
Because the shifts might be non-linear, the shifts $\Delta x_i$ won't be pure numbers but something more like shift operators.
That is, if there were a true spectrum $\tilde{y}$, each individual noisy image $y_i$ would be related to that true spectrum by a shift operator acting on that true spectrum, plus noise.
We will assume here that (from some external information) we know these shift operators very accurately.
This is often not true, of course; we will return to this point in \secref{sec:discussion} below.

\paragraph{Wavelength calibration}
We will assume that, for every $M_i$-pixel image $y_i$ we also have a $M_i$-vector (or list) $x_i$ of pixel positions, such that each element of $x_i$ gives an accurate and precise value for the wavelength (or position in the spectrograph, in some settings) for each pixel of the image $y_i$.
That is, the device is well calibrated, and we know all the housekeeping data we need for image $y_i$.

If the data being combined were images rather than spectra, the relevant calibration would be astrometric calibration; to wit, that each pixel has a known position on the sky.

\paragraph{Noise model}
Each observation is noisy.
We will assume that the noise has good properties, and especially that it is additive, (nearly) zero-mean and (nearly) Gaussian in form.
We will \emph{not} assume that you know the variance of the noise accurately, but you can do better with this method if you do know the noise variances accurately for each pixel.
In what follows, each spectrum $y_i$ is given an associated (but possibly inaccurate) noise-variance tensor $C_i$.
If $y_i$ is an $M_i$-vector, then $C_i$ must be a nonnegative-definite $M_i\times m_i$ 2-tensor.
In this description, there is no need to assume that the different pixels of the spectrum received independent noise, but we do assume that each epoch $i$ has an independent noise draw.

\paragraph{Bad-pixel masks}
Occasionally (or frequently) some of the pixels of a spectrum $y_i$ might be bad---affected by cosmic rays or electronics issues---such that there is a bad-pixel mask $b_i$ associated with each spectrum $y_i$.
This mask has value 1 in the locations in all good pixels, and 0 in the locations of all bad pixels.
There are many different ways in which a pixel can be bad; in what follows the assumption is that the relevant judgements have been made more-or-less correctly.

\paragraph{Data gaps}
There is no assumption here that all the $y_i$ have the same number of pixels $M_i$, nor that the pixel grid is in any sense uniform or complete.
That is, there can be gaps and spaces in the wavelength coverage of the spectra, as there are at chip gaps and where the detector has bad columns.

\paragraph{Pixel-convolved line-spread function}
The spectrograph has something like a point-spread function or line-spread function, which, in the case of spectroscopy, sets the shape of an unresolved spectral line.
In general it depends on the slit or fiber width, the spectrograph resolution and optics, instrument focus, and the properties of the detector.
In what follows we will not build a model of any of this.
Instead we will assume that there is a relatively constant pixel-convolved line-spread function (PCLSF), constant both in time, and constant with respect to the shifts $\Delta x_i$ we see in our data set.
We consider only the PCLSF, because then the sampling by the detector pixels does not require any additional convolution.
That is, when the PCLSF is used to model a spectrum, the projection onto the pixels is just a sampling of the function at the pixel centers, and does not involve any convolution over the face of the pixel.
All that said, we won't explicitly model or use the PCLSF in what follows; we will just assume that our continuous spectral model is a model for the pixel-convolved, finite-resolution spectrum; we won't be doing any convolutions in comparing models and data.

In many cases the LSF or PSF is actually substantially different between exposures.
This leads to spectral (or image) variations, and violate the assumptions of this model (and also the implicit assumptions underlying standard practice).
We won't address this kind of variability here, except for some comments in \secref{sec:discussion}:
The method presented here can be straightforwardly modified to account for LSF or PSF variations.

\paragraph{Continuous spectral model} 
In what follows we are going to avoid interpolating the data.
We can avoid it by having a spectral \emph{model} instead that can be arbitrarily and losslessly interpolated.
That means that the spectral model must be a representation of a continuous one-dimensional function.
If we were working on two-dimensional images, the model would have to be a representation of a continous two-dimensional function.

\paragraph{Linear basis or mixture models}
The simplest kind of model for a continuous function is a sum or mixture or linear combination of continuous basis functions.
These could be sines and cosines for a Fourier series model.
They could be spline (or sinc or Lanczos) cardinal basis functions.
They could be wavelets.
The point is that if we use a linear basis, and assume that noises are Gaussian, we will get closed-form expressions for the data combination.
Thus we will make use of a linear basis.

\paragraph{Band limit}
We are not assuming, in what follows, that the result we are looking for will be band-limited;
we say that a signal is \emph{band-limited} if it has no frequency content above some highest possible cutoff frequency.
There is a weird sense in which data from a spectrograph both is, and is not, band limited.
It is band limited in that the spectrograph is finite in resolution, such that no real spectroscopic signal can show features narrower than the PCLSF.
It is not band-limited in that the photon noise in the device is independent (or close to independent) from pixel to pixel, such that the noise contribution to each observation $y_i$ has support at all spatial frequencies.

One interesting question in what follows is whether the forward model we build for the data should itself be restricted to the band limit defined by the spectrograph resolution, or whether it should permit higher frequencies.
It might surprise the reader to learn that we prefer to let the model capture higher frequencies, even when we think they shouldn't be there for any reasons other than noise, given the hardware.
The reason we provide our model the flexibility to go to higher frequencies is that we want our combined spectrum to have the capacity to capture or represent the pixel-to-pixel noise in the estimate; this pixel-to-pixel noise obeys no band limit.

\paragraph{Critical sampling}
We are not assuming, in what follows, that the data are critically sampled.
Because of considerations related to the Nyquist frequency on a uniform grid, it is valuable for spectra to be represented with a pixel spacing that is smaller than the width of the PCLSF.
Specifically, it is valuable for there to be two or three pixels across the full-width at half-maximum (FWHM) of the PCLSF.
This keeps the band-limited part of the spectroscopic signal well-sampled in the data.
In general, interpolation works much better on critically sampled data than it does on ill-sampled data.

The fact that we are not interpolating the data means that we don't require critical sampling in the raw data.
And indeed, there are spectrographs that don't have critically sampled pixel scales (for example, \textsl{APOGEE}; \citealt{apogee}).
Thus we will not assume that the raw data are critically sampled, but we will produce the output combined spectrum on a grid that is critically sampled with respect to the instrument resolution.
The input and output pixel grids do not need to coincide when the model is a continuous function.

\paragraph{Non-uniform fast Fourier transform}
In the special case that the spectral model is a Fourier series, there are very useful and fast tools for interpolation, including especially the non-uniform fast Fourier transform (for example, \citealt{finufft}).
These tools make the spectral combination substantially faster than naive linear algebra estimates would suggest.

\paragraph{Time variability}
When we average data (or build a static model of multi-epoch data), we are implicitly assuming that we do not care about any of the intrinsic variability of the source being observed.
That is, we are assuming that the changes from observation to observation are solely from either the shifts $\Delta x_i$ or else from effects that we consider to be noise.
We interpolate the model (not the data!) to account for the shifts, and we combine data to average down this noise.
In many cases the stellar variability is itself the subject of study, and then the variability is not entirely noise.
But from the perspective of this project, any stellar variability is considered to be part of the total additive noise in the spectrum.
We return to this subject below in \secref{sec:discussion}.

\paragraph{Bias, sky, tellurics, flat-field, normalization}
In what follows, we will assume that the input data are well calibrated in various ways.
In particular, we will assume that the images or spectra have been bias-corrected, sky-subtracted, flat-fielded, and tellurics-corrected sufficiently well that the different observations $y_i$ can be seen as observations of the same underlying or intrinsic signal.
Alternatively, the input data might be continuum-normalized or pseudo-continuum-normalized.
Again, the assumption is that the different calibrated or processed observations are of the same underlying signal.

Many of these effects can be accounted for in straightforward generalizations of what we deliver below.
For example, if there are multiple signals participating, the model would have to expand into something like \textsl{wobble} (\citealt{wobble}).
For another, if there are normalization variations, a normalization model can be added to the image combination procedure.
That is, if different input spectra have different normalizations, the model for the combined spectrum can be multiplied by a normalization model, unique to each input spectrum, prior to evaluation of the likelihood.

\paragraph{Flux or photon conservation}
Some methods for image combination, like \texttt{Drizzle} \citep{drizzle}, have as a hard constraint that they conserve \emph{flux} or \emph{photons}.
We do not agree with this constraint or condition.
The fundamental observable in a telescope is intensity, or energy per area per solid angle per time per wavelength (or per frequency).
This quantity is related to the phase-space density of photons.
When the mapping between a pixel grid and the sky is changed, the expectation of the rate of photon arrivals per pixel will change (because the pixel effective solid angles change), but the expectation of the intensity will not change.
Our position is that the key idea is not the conservation of flux or photons, but that the spectra or images being combined make the same measurements of the intensity (or similarly normalized or otherwise consistently calibrated measurements of intensity) at the detector surface.
The combined image should provide a good mean estimate of this intensity.

\section{Method: \name}\label{sec:method}

The problem set-up is that there are $N$ spectral measurements $y_i$ (with $1\leq i\leq N$), each of which is a $M_i$-pixel list of intensities (fluxes maybe).
Different observations $i$ might have different numbers $M_i$ of pixels.
The $M_i$ pixels are associated with an $M_i$-length list of positions (wavelengths or log-wavelengths) $x_i$.
Each measurement or observation $i$ is shifted by a shift operator $\Delta x_i$, which can be thought of as a length-$M_i$ list of pixel offsets or shifts, such that the ``rest frame'' or ``unshifted'' positions corresponding to the length-$M_i$ list of data $y_i$ is the list of differences $x_i - \Delta x_i$.
The goal is to get a combined spectrum that represents the mean of these spectra, in the common rest frame.

The spectral expectation model $f(x;\theta)$, which will be a representation of our combined spectrum, is a function of $x$ controlled by a list $\theta$ of $P$ linear parameters $a_j$ (with $1\leq j\leq P$).
This function will be very flexible; possibly even an interpolation of control points or a Fourier series; the number $P$ will be roughly the number of pixels we want in our output combined spectrum.
The statistical model underlying can be expressed qualitatively as
\begin{align}
    y_i &= f(x_i - \Delta x_i;\theta) + \mbox{noise} \\
    f(x;\theta) &= \sum_{j=1}^P a_j\,g_j(x) ~,
\end{align}
where we are being a bit loose with terminology (since $y_i$ is a set of fluxes, and $x_i$ is a set of positions, not just one position), the noise is additive (but not yet specified), and each of the $P$ functions $g_j(x)$ is a basis function of the representation.

The noise model is additive (as noted above) and we will assume that the noise contributing to observation $y_i$ is zero-mean and Gaussian, with a known $M_i\times M_i$ variance tensor $C_i$ (or its inverse $C_i^{-1}$).
If the raw pixel measurements are independent, which they often are (and are often assumed to be), then the $C_i$ (or $C_i^{-1}$) matrices will be diagonal matrices with the individual pixel variances (or inverse variances) down the diagonals.
In many cases you don't know (or aren't provided with) the individual pixel variances, and you are just co-adding data with uniform weights.
That choice, for our purposes, is equivalent to setting every inverse variance matrix $C_i^{-1}$ to the $M_i\times M_i$ identity matrix.
That's permitted!
It is no worse a choice than it was (in the past) when you averaged your input data without thinking about noise variances or weights.
Another sensible default choice would be to set each inverse matrix $C_i^{-1}$ to the identity times the exposure time, such that the eventual math we do will weight the data by the exposure times.
That is, you do not need to know the noise variances accurately (or even at all) to execute the forthcoming procedure; but if you do know them, it is good to use them.

If you additionally have a bad-pixel mask $b_i$ associated with each observation $y_i$, the bad-pixel mask can be used simply to zero out the corresponding bad rows and columns of the inverse covariance matrix $C_i^{-1}$.
Importantly, the bad data get zeros in the \emph{inverse} covariance matrix, not the covariance matrix!

Because we are often dealing with a spectrograph with a (fairly) well defined spectral resolution ($\delta\lambda/\lambda$ roughly constant), and because we care a lot about Doppler shifts (which are uniform in log wavelength), it makes sense to think of the pixel positions $x_i$ and shifts $\Delta x_i$ as being in log-wavelength (or ln-wavelength) units.
Indeed, the \textsl{Sloan Digital Sky Survey} family of spectrographs extract one-dimensional spectra on a uniform-in-log-wavelength basis.
Once that choice is made, there are still many choices for the linear basis,
which include canonical functions from interpolation bases, wavelets, and Fourier series.
In what follows, somewhat arbitrarily, we will use a Fourier basis, which can be expressed as follows:
\begin{align}
    g_j(x) & = \left\{\begin{array}{cl}\displaystyle\cos\left(\frac{\pi\,[j-1]}{L}\,x\right) & \mbox{for $j$ odd} \\[3ex]
    \displaystyle\sin\left(\frac{\pi\,j}{L}\,x\right) & \mbox{for $j$ even}\end{array}\right. ~,\label{eq:basis}
\end{align}
where $L$ is a (long) length-scale in the $x$ space (the log-wavelength space), and $1\leq j\leq P$.

Now for the method:
We append all of the observations $y_i$ into one huge $M$-pixel list $Y$, where $M=\sum_{i=1}^N M_i$.
We make a $M\times M$ block-diagonal matrix $C^{-1}$ from all the individual inverse variance matrices $C_i^{-1}$.
We make a $M\times P$ design matrix $X$ which is the evaluation of all the $P$ basis functions at all the pixel locations in all of the $x_i$.
Once we have these things, the best-fit parameters $\hat\theta$ are
\begin{align}
    \hat\theta &= (X^\top\,C^{-1}\,X)^{-1}\,X^\top\,C^{-1}\,Y ~, \label{eq:xtxinvxty}
\end{align}
which is the maximum-likelihood or minimum-$\chi^2$ solution for a linear weighted least squares problem or a linear model with Gaussian noise (see, for example, \citealt{fitting}).
And now if we choose a final (uniform, say) output pixel grid $x_\star$ in the rest-frame (unshifted) wavelength space, the combined spectrum is just
\begin{align}
    y_\star &= X_\star\,\hat\theta ~,
\end{align}
where $X_\star$ is the $M_\star\times P$ evaluation of all of the $P$ basis functions at the locations of the $M_\star$ pixels in the output combined-spectrum pixel grid $x_\star$.
Putting these things together, the output spectrum $y_\star$ can be thought of as just a kind of linear weighted averaging operation on the input spectra $y_i$ (which have been packed into $Y$) according to
\begin{align}
    y_\star &= A_\star\,Y \\
    A_\star &\equiv X_\star\,(X^\top\,C^{-1}\,X)^{-1}\,X^\top\,C^{-1} ~.
\end{align}
That is, the method we are delivering is a linear combination of the input data, just like traditional interpolate-and-average.
But our method never involves, not even implicitly, shifting or interpolating the input data.

The uncertainty on the combined spectrum $y_\star$ can be estimated in a few ways.
The information-theoretic answer is that the covariance matrix $C_\star$ delivering the uncertainties on $y_\star$ is
\begin{align}
    C_\star &= [X_\star\,(X^\top\,C^{-1}\,X)^{-1}\,X_\star^\top]^{-1} ~.\label{eq:outcovar}
\end{align}
The naive uncertainty variances are the diagonals of this matrix (and not the inverses of the diagonals of its inverse).
But, \emph{very importantly}, this is only a correct usage when the input $C_i$ matrices really represented the true uncertainties going in.
If you have any concerns about that, then it is way safer to get your uncertainty estimates from bootstrap or jackknife (for a related example, see \citealt{fittingflexible}).

In the end, most users want the combined spectrum $y_\star$ (and perhaps the pixel positions $x_\star$, uncertainty information, and propagated meta data).
Some users will additionally want (or want to reconstruct) the parameters $\hat\theta$ and the basis functions $g_j(x)$.
For these reasons it often makes sense to set $M_\star\geq P$ such that the delivered $y_\star$ fully determines the parameters.
But that's a detailed question about your users and your goals.

\paragraph{Implementation notes}
Although the equations like \eqref{eq:xtxinvxty} show matrix multiplies between $C^{-1}$ and $X$, it is not a good idea to perform these multplies as true matrix multiplies if the $C^{-1}$ matrix is diagonal (as it often is).
When $C^{-1}$ is diagonal, the matrix multiply can be done as a broadcast of the vector of diagonal values instead, which is faster (and way less memory intensive).

Although the equations like \eqref{eq:xtxinvxty} show inverses, it is almost never a good idea to use the native linear-algebra inverse function \texttt{inv()} in any implementation of this method.
The reason is that we don't want the inverse matrix itself, we want the outcome of the inverse matrix multiplied by $X^\top\,C^{-1}\,Y$.
For this use case, it is more precise and more more numerically stable to use the linear algebra \texttt{solve()} or \texttt{lstsq()} functions.
The only use for \texttt{inv()} is where the full inverse of a matrix is required, as when estimating the full $C_\star$ by \eqref{eq:outcovar}.

Whenever performing \texttt{inv()}, \texttt{solve()}, or \texttt{lstsq()} operations, it is sensible to begin by checking the condition number (the ratio of the largest to the smallest singular value) of the matrix in play.
Condition numbers larger than $10^8$ are usually trouble, even when the mathematics is being performed at double precision.
When condition numbers get large, maybe a different basis or a re-scale of some basis functions might be a good idea.

If $X$ is either very sparse (as it is in a b-spline basis) or Fourier (as it is in a sine-and-cosine basis) then there are very fast ways to do the linear algebra.
We could have used the \texttt{finufft} package (\citealt{finufft}) to do the mathematics fast, because we have a Fourier basis here.
Sparse linear algebra packages like that in \texttt{scipy} (\citealt{scipy}) are available; they would similarly speed up the code when using spline-like or wavelet-like bases.

If you use any non-uniform fast Fourier transform package, the programming interface of that package will constrain how the frequencies must be ordered in the code, and it will probably require complex-number implementation.
These have impacts on how the code is written.
In particular the fact that the output combined spectrum $y_\star$ is real-valued constrains the complex Fourier amplitudes.
Continuing on the Fourier theme:
Fourier series, technically, model \emph{periodic} functions.
The length scale $L$ of that periodicity in \eqref{eq:basis} must be set to a length longer than the $x$-space span of any of the input or output pixel grids.

We recommended setting $P=M_\star$.
We recommend very strongly \emph{against} setting $P\approx M$, where $M=\sum_i M_i$.
That is, you want the number of parameters to equal your number of output pixels, but you \emph{don't} want your number of parameters to be close to the number of \emph{input} pixels.
The reason for this is the ``double descent'' phenomenon, which de-stabilizes linear fitting (\citealt{doubledescent}).
If you find yourself in the situation that $M_\star\approx M$, reconsider your life choices.

\section{Experiments and results}\label{sec:results}

\paragraph{Fake data}
We demonstrate the effectiveness of \name{}---the spectral combination method by forward modeling---on artificial data.
The spectral data were generated from a mean spectral expectation that is a continuum plus a randomly generated line list, where each line equivalent width was set by a random draw from a power law. 
This line formation model is unphysical, but it is sufficient for our purposes in that it generates fake data that look like stellar spectra.
Each spectral epoch was given a velocity (Doppler shift) by putting the epochs equally spaced on a sinusoidal variation with a semi-amplitude of $30\,\kmps$ (to emulate the motion of the spectrograph with respect to the Solar System barycenter).
The spectral expectation model is built on the assumption of a normalized (unit) continuum and unresolved spectral lines with a Gaussian line-spread function (LSF) with one-sigma width $1/R$, where $R=135\,000$ is the spectrograph resolution.

The sampled data at each epoch were made by simply evaluating the spectral expectation model (shifted by the Doppler shift) at the pixel centers, such that the $R=135\,000$ LSF is effectively the pixel-convolved LSF.
After sampling the mean spectral expectation model onto an observational pixel grid, Gaussian noise is added with a variance that grows linearly with expected flux, such that the signal-to-noise per pixel in the continuum has a definite known value in each epoch spectrum.
At a rate of 0.01 (one percent), sets of three adjacent pixels were randomly and independently marked as ``bad'' and offset by a large positive offset.

\begin{figure}[t!]
    \begin{mdframed}\begin{center}
    \includegraphics[width=1.3\figurewidth]{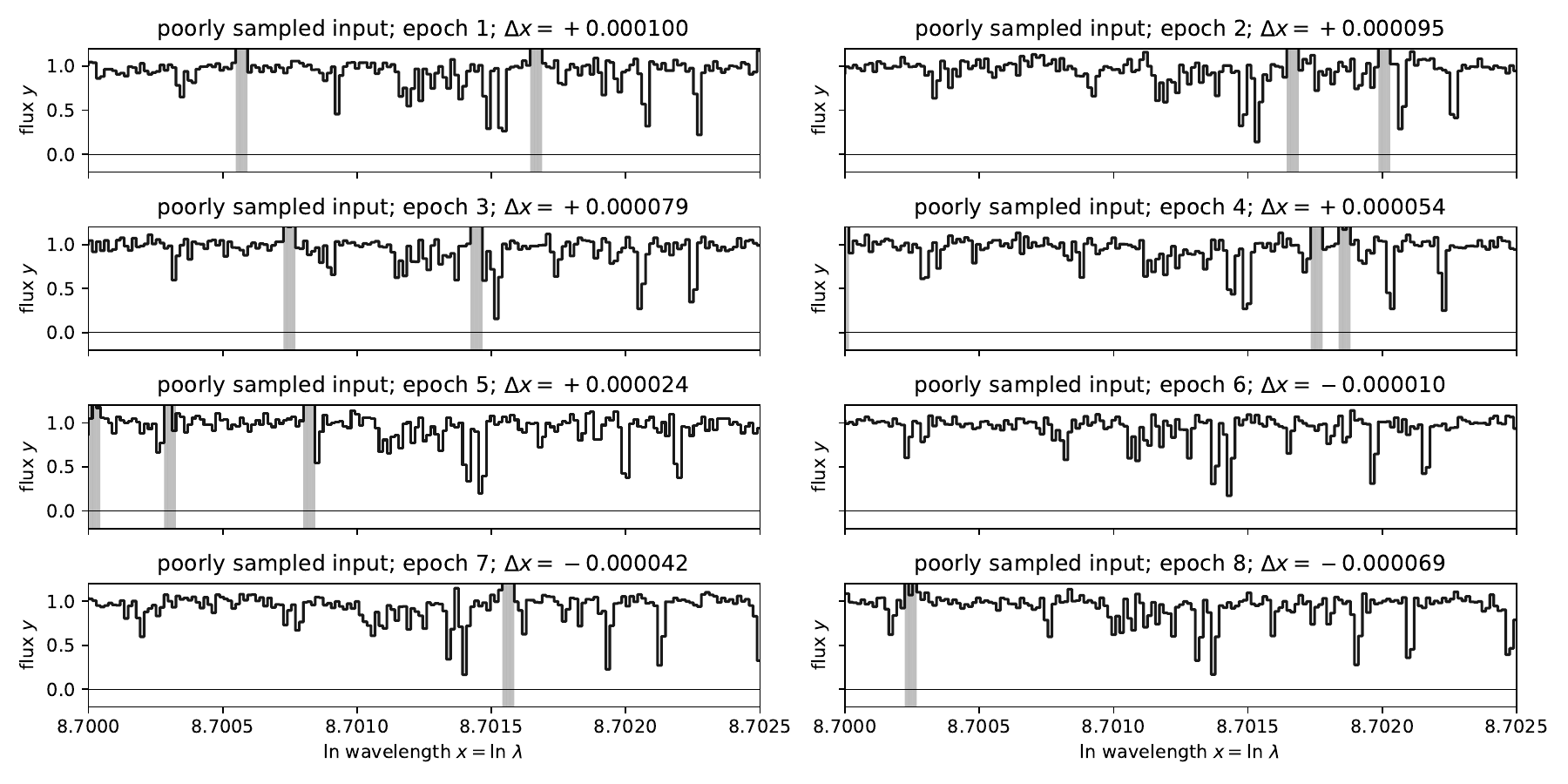}\\
    \includegraphics[width=\figurewidth]{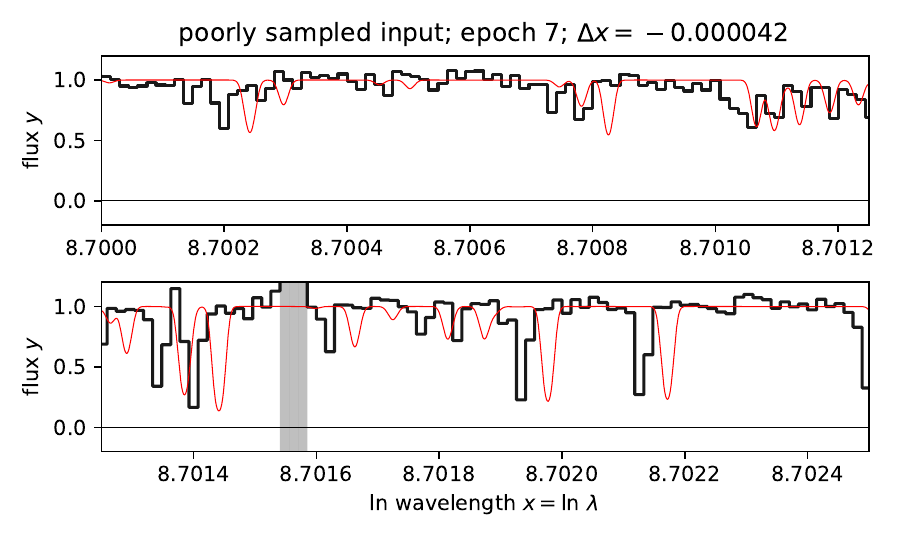}
    \end{center}
    \caption{The top 8 panels show the $N=8$ epochs of the poorly sampled input data, with $N=8$, $M=171$, and signal-to-noise per pixel of 18 in the continuum. The bad pixels are indicated with grey bars. The bottom 2 panels show a zoom in on one of the epochs, along with the true spectrum used to generate the data. The true spectrum is shown at zero Doppler shift, while each epoch spectrum is at a finite Doppler shift $\Delta x_i$ relative to the spectrograph pixel grid.\label{fig:data1}}
    \end{mdframed}
\end{figure}

\begin{figure}[t!]
    \begin{mdframed}\begin{center}
    \includegraphics[width=1.3\figurewidth]{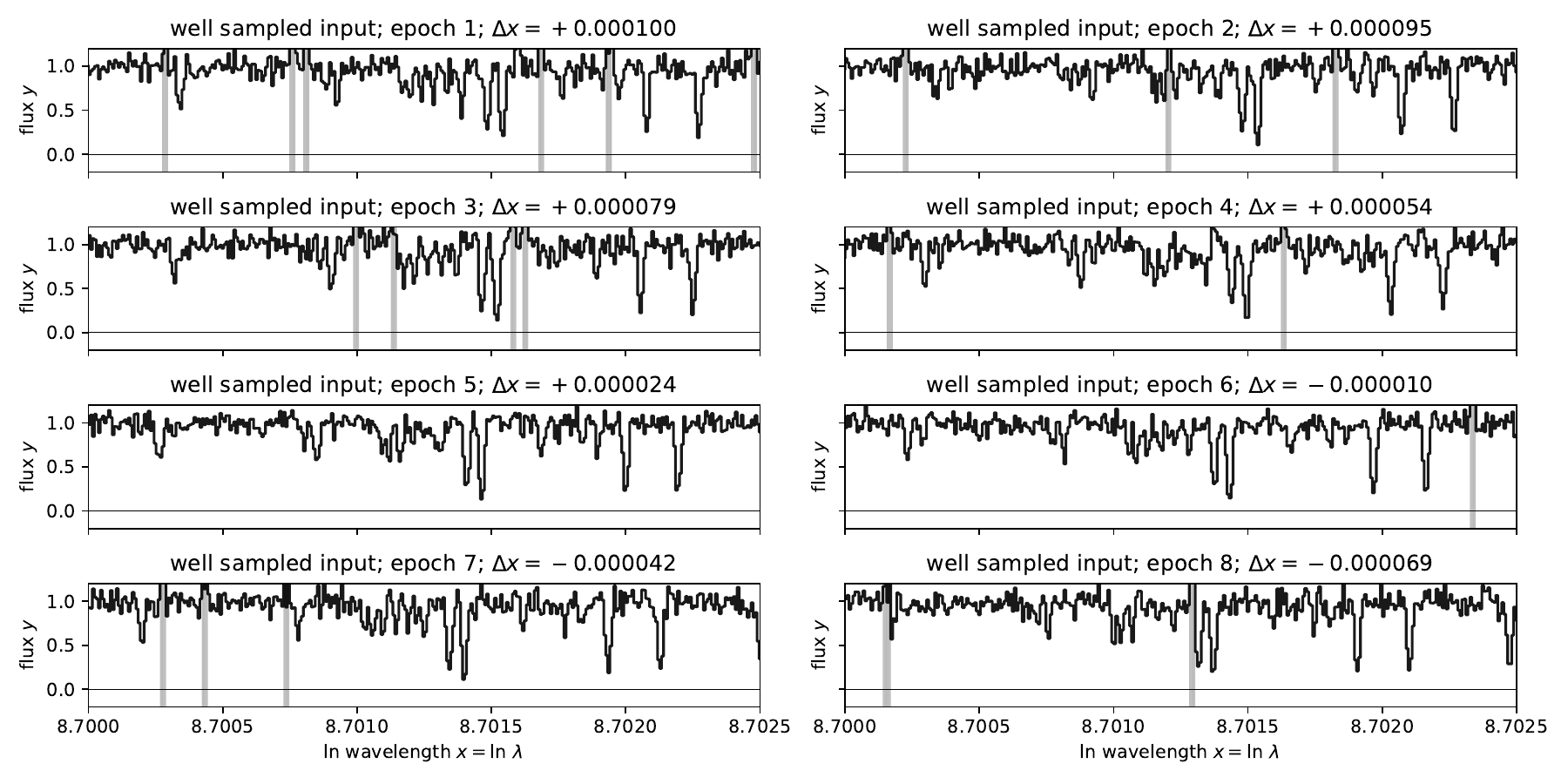}
    \end{center}
    \caption{Similar to the top panels of \figurename~\ref{fig:data1} except showing the well sampled input data with $N=8$, $M=340$, and signal-to-noise per pixel of 12 in the continuum.\label{fig:data2}}
    \end{mdframed}
\end{figure}

According to these rules we made two input data sets.
One is poorly sampled, with $M=171$ spectral pixels in the input data separated by $2/R$ in $x$ (natural logarithm of wavelength), and signal-to-noise per pixel of 18 in the continuum.
The other is well sampled, with $M=340$ spectral pixels in the input data separated by $1/R$ in $x$, and signal-to-noise per pixel of 12 in the continuum.
In each case there are $N=8$ epochs.
The two complete data sets are shown in \figurename~\ref{fig:data1} and \figurename~\ref{fig:data2}.

\paragraph{\name{}}
As our main experiment or result, we apply the forward model described above in \secref{sec:method} to the input data shown in \figurename~\ref{fig:data1} and \figurename~\ref{fig:data2}.
In each case, the output pixel grid $x_\star$ is chosen to be well sampled, with a pixel spacing of $1/R$ in the (natural) logarithm of wavelength.
In order to give the model maximum flexibility, the number $P$ of Fourier modes in the spectral model is set to be equal to the number of pixels $M_\star$ in the pixel representation.
The log-wavelength scale $L$ of the Fourier basis was set to the pixel spacing times the number of pixels.
We didn't employ any data weighting; we treated every epoch as identical in terms of inverse variance.

\begin{figure}[t!]
    \begin{mdframed}\begin{center}
    \includegraphics[width=\figurewidth]{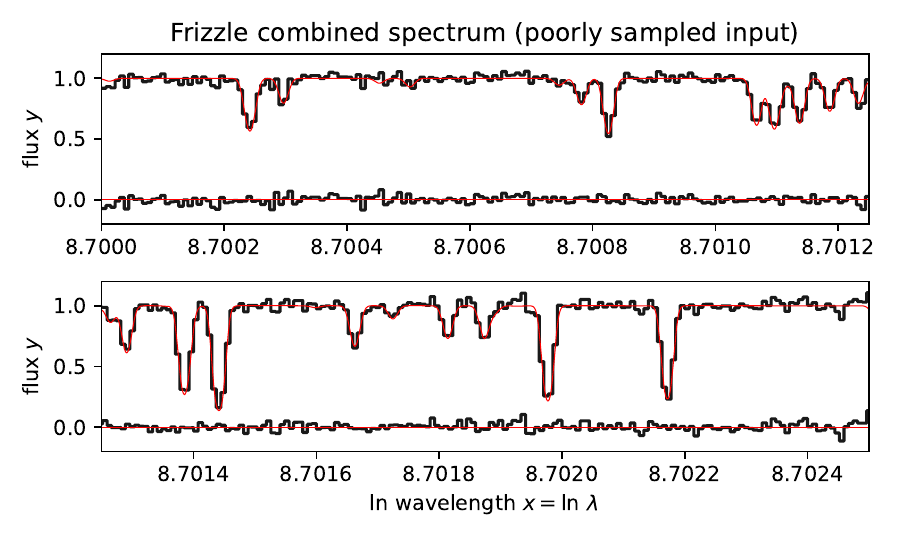}\\
    \includegraphics[width=\figurewidth]{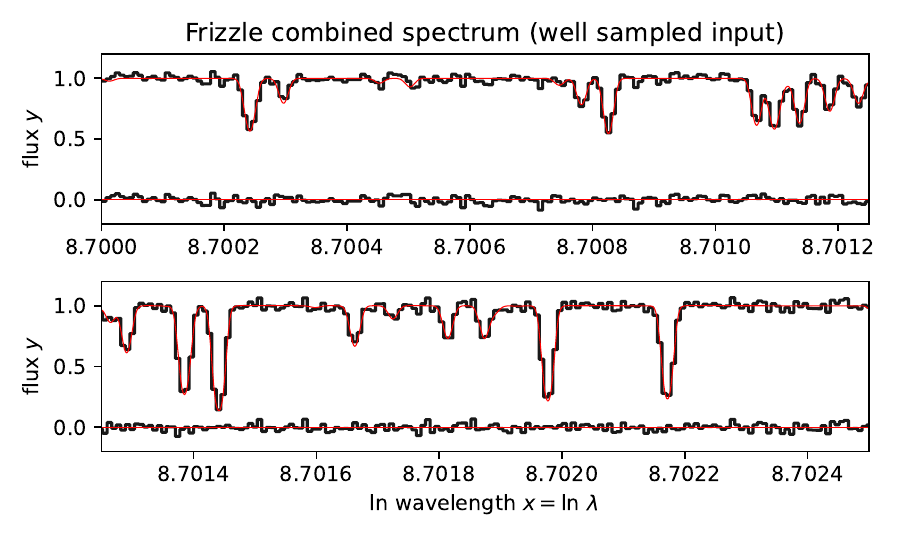}
    \end{center}
    \caption{The top two panels show the result of running \name{} on the poorly sampled input data, plus the residual away from the truth. The bottom two panels show the same but for the well sampled input data. In each panel, the red line shows the true model used to generate the data.\label{fig:forward}}
    \end{mdframed}
\end{figure}
The combined spectrum resulting from the forward model described in \secref{sec:method} is shown in \figurename~\ref{fig:forward}.
It is compared to the true spectral expectation model that was used to generate the data.
Also shown in \figurename~\ref{fig:forward} is the residuals, \name{} minus truth.
The residuals look stationary, approximately Gaussian, and uncorrelated (more on this below).

\paragraph{Standard Practice(tm)}
In standard practice, the individual spectra $y_i$ are interpolated to rest-frame spectra $y'_i$ using an interpolator.
In detail, the data are shifted and also resampled onto the output pixel grid $x_\star$.
The interpolator is a choice; cubic spline, Lanczos, and sinc interpolations are common, since they have good properties with respect to the spectrograph band limit.
Here we choose a cubic spline for convenience, but there are many other choices, including linear, sinc, and Lanczos interpolators, among many others.

The interpolated spectra $y'_i$ are, by construction, all on the same rest-frame wavelength grid $x_\star$ so they can be averaged, with a straight average, a median, an inverse-variance-weighted mean, or any more sophisticated algorithm.
Here we choose an unweighted mean, with the exception that we zero out the contributions of any bad pixels in that mean.

In Standard Practice(tm), this censorship of the bad pixels requires some attention:
Naive interpolation of an integer (binary actually) bad-pixel mask $b_i$ will not deliver an integer mask $b'_i$.
The conservative move (which we adopt) is to naively interpolate $b_i$ (also by the cubic spline) and then zero out any pixels that are significantly below unity.
In general this substantially grows the binary mask, and reduces the amount of data available for combination.
The non-conservative approach is just as bad: if we restrict the number (or fraction) of pixels in the combined spectrum to equal those in the individual epochs, we risk un-masking genuinely bad pixels while growing the binary mask elsewhere!

\begin{figure}[t!]
    \begin{mdframed}\begin{center}
    \includegraphics[width=\figurewidth]{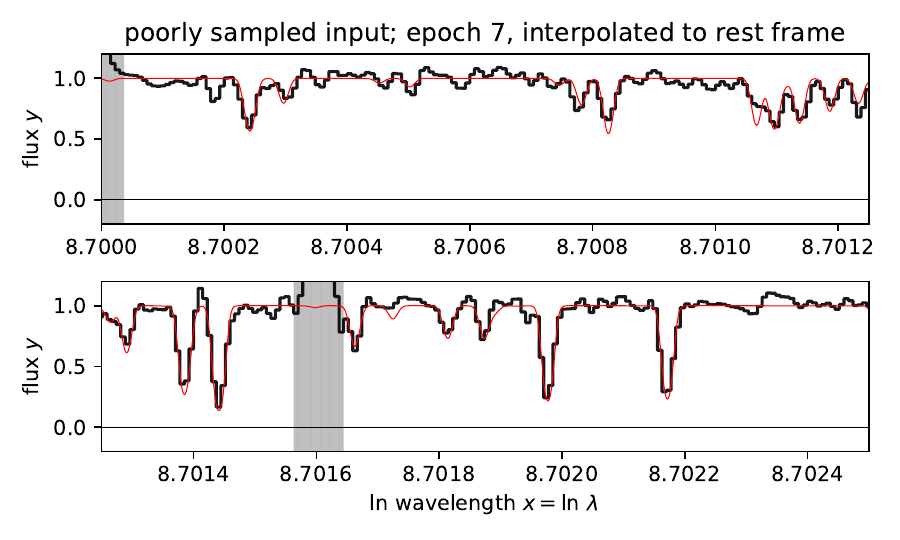}
    \end{center}
    \caption{As part of Standard Practice(tm), it is necessary to interpolate the individual epoch spectra before averaging them. This shows one epoch of the poorly sampled data, interpolated to the output $x_\star$ pixel grid in the rest frame. The interpolated spectrum is compared to the true spectral expectation employed to generate it. Note the ringing induced by the interpolation. Also shown is the censored data, near an interpolated bad pixel and at the edge of the domain. Compare to the bottom panels of \figurename~\ref{fig:data1}.\label{fig:interpolated}}
    \end{mdframed}
\end{figure}
An example of a poorly-sampled spectrum interpolated to the rest frame is shown in \figurename~\ref{fig:interpolated}.
The interpolation introduces substantial ringing artifacts in the spectrum (compare to \figurename~\ref{fig:data1}).
It also grows the bad-pixel mask around the bad pixel.

\begin{figure}[t!]
    \begin{mdframed}\begin{center}
    \includegraphics[width=\figurewidth]{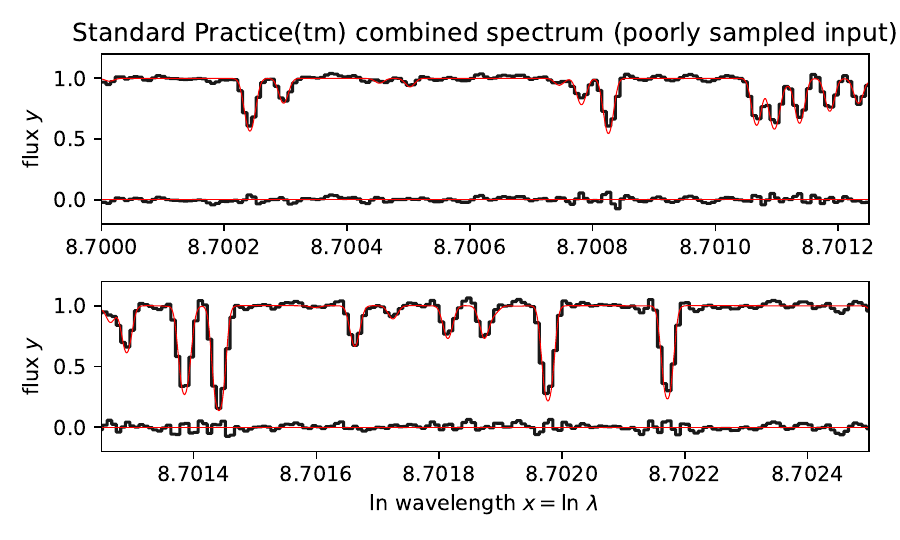}\\
    \includegraphics[width=\figurewidth]{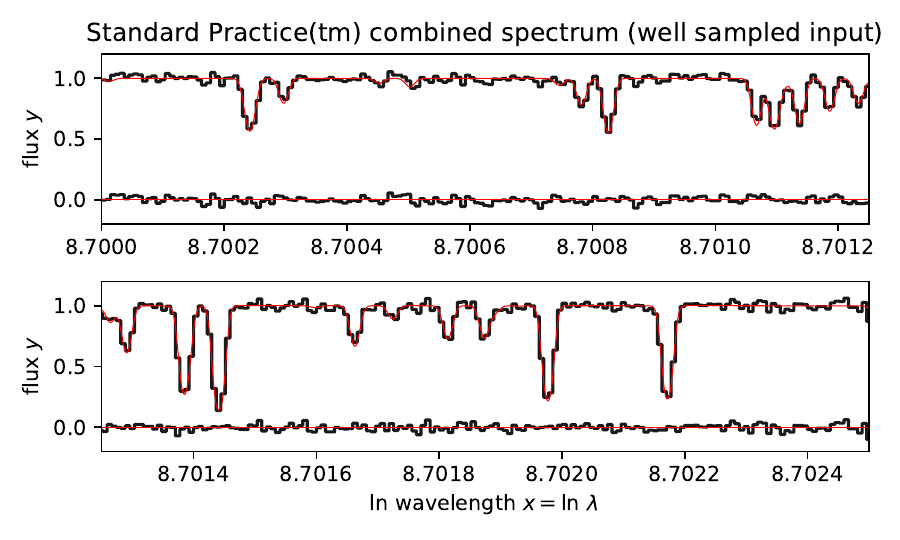}
    \end{center}
    \caption{Same as \figurename~\ref{fig:forward} but for the output of Standard Practice(tm). Note that in the poorly sampled case, the residuals appear to be spatially correlated, and the residuals appear to be larger near strong spectral features; more about these issues in the text and in \figurename~\ref{fig:noise}.\label{fig:standard}}
    \end{mdframed}
\end{figure}
We show the final result of Standard Practice(tm)---interpolating the input data and then averaging the interpolations---for the two data sets, in \figurename~\ref{fig:standard}.
We also show the comparison with the true spectral expectation model that was employed to make the fake data.
The residuals look correlated (there is ringing which mimics weak spectral features), and the residuals are larger near strong spectral features, especially in the case of the poorly sampled input data.

\begin{figure}[t!]
    \begin{mdframed}\begin{center}
    \includegraphics[width=\figurewidth]{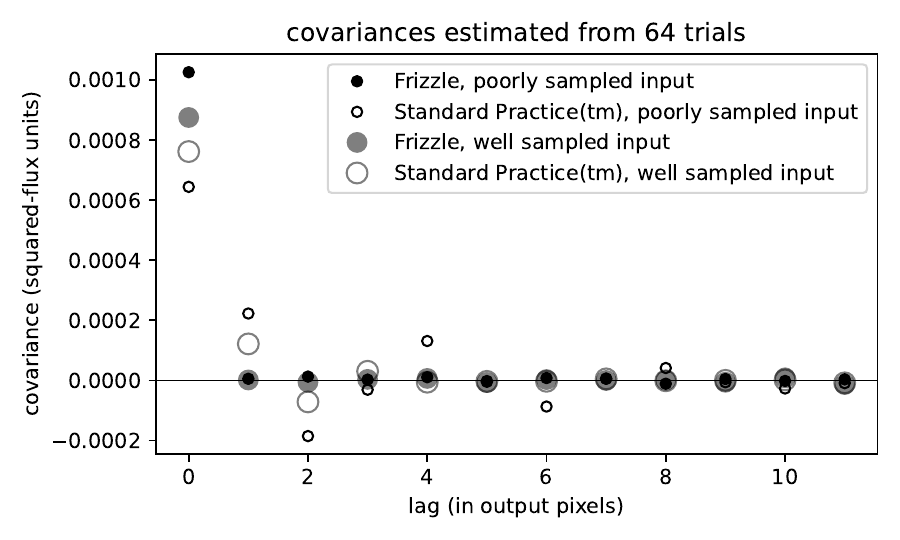}
    \end{center}
    \caption{The empirical noise variance in the output (combined) spectra, and pixel-to-pixel covariances, as a function of pixel lag (pixel offset). The variances were estimated by performing the experiments repeated 64 times, with the same spectral expectations and sampling but unique noise draws. \name{} has slightly larger variance per pixel at zero lag, but vanishing pixel-to-pixel correlations. In contrast, the Standard Practice(tm) leads to correlated pixel values, with correlations extending out to many pixels.\label{fig:noise}}
    \end{mdframed}
\end{figure}

\paragraph{Noise and noise covariances}
We repeated the above experiments to empirically measure the noise covariances, and to estimate how correlated neighboring pixels are in the output spectrum. For the poorly sampled case and the well sampled case, we generated 64 multi-epoch data sets and computed the mean spectrum using \name{} and Standard Practice(tm). We then computed the empirical covariance in the output spectrum minus the true spectrum, which are shown in \figurename~\ref{fig:noise}.
In each case, Standard Practice(tm) has a lower noise variance at zero lag, but shows spatially correlated noise over many pixels.
\name{}, in contrast, shows no pixel-to-pixel noise covariances, even when the input data are ill sampled.

It is interesting to ask \emph{why} the output combined spectrum in the forward-modeling case shows uncorrelated noise.
It won't in general: If the number of epochs is small, or the coverage in the wavelength domain is spotty, there will in general be pixel-to-pixel covariances in the uncertainties on the combined spectrum.
However, if the number of epochs is large and the distribution of shift operators $\Delta x_i$ is good, as it is in these experiments, the off-diagonal elements of the output combined-spectrum covariance matrix given in \eqref{eq:outcovar} will tend towards zero.
The Standard Practice(tm) output combined spectrum always shows correlated noise, and especially when the input data are ill-sampled.
The reason for this is that interpolation is very sensitive to noise in the ill-sampled case, such that the inter-pixel value of the interpolated function is moved substantially by small movements in the input pixel values, and these inter-pixel regions in the raw data are large relative to the pixel spacing in the combined spectrum spectral grid.

\section{Discussion}\label{sec:discussion}

We have presented a forward-modeling approach to combining spectra or images;
we name it \name.
\name{} has good properties:
It doesn't involve interpolating or otherwise distorting the input data;
it produces a combined or mean spectrum in the format that astronomers traditionally expect;
it makes no assumptions about the form of the combined spectrum (except that it can be represented on an $M_\star$-pixel grid);
and it has excellent noise properties in real circumstances.
The method can work on a wide variety of different kinds of data, but it is especially valuable in projects where the individual input exposures are ill-sampled, or not as well sampled as one might like; our original motivation was the individual \textsl{APOGEE} spectrograph (\citealt{apogeehardware}) exposures.
It is also good for situations in which the input data have heterogeneous bad pixels or data gaps.

The method does not involve interpolating the data.
However, it does fit a very flexible model to the data, and there is a limit in which fitting very flexible models is equivalent to interpolation (see, eg, \citealt{fittingflexible}).
So is this method, implicitly, interpolating the data?
Our answer is no.
Interpolation of the data involves non-parametric models with control points that are set by the pixel grids of the input data.
Our model choice never refers to the properties of the input data pixel grids; it is chosen based on the desired properties of the output, combined-spectrum grid.
Furthermore, interpolation functions (like cubic spline, say, or Lanczos-5, say) pass through every raw-data pixel value exactly.
There is no point in this method at which the function being fit passes through every (or even any) of the raw-data pixel values.

In the Introduction (\secref{sec:intro}), we were agnostic about whether we were talking about imaging or spectroscopic data.
In the end, our examples are spectroscopic.
What differences are there between these cases?
The main difference is that there are many more ``shift'' parameters for imaging than spectroscopy:
Two spectra (according to our assumptions) differ only (or primarily) in terms of a one-dimensional Doppler shift with respect to pixels that are located in one dimension (the wavelength direction).
Imaging epochs, on the other hand, can differ in terms of two shifts and a rotation, or (equivalently) three Euler angles.
And, since every pixel has its own two-dimensional position, the problem of locating pixels in an image is generally harder than the problem of locating pixels in an extracted spectrum.
That is, the imaging case is generally more difficult.
But it is not different, conceptually, from what is presented above.
That is, once the (now more complicated) shift operators $\Delta x_i$ are known, the problem proceeds by the same mathematics.

Everything presented above depends on a set of strong assumptions, listed in \secref{sec:assumptions}.
Perhaps the strongest assumption is that the spectrum or image is not a function of time.
That is, each epoch spectrum or image is a sampling of the same underlying, true signal.
This assumption can be wrong for a myriad of reasons:
The source can vary, the instrument can vary, there can be variable foregrounds or backgrounds, and there can be mistakes or variations in calibration or data processing.
All of these things happen in real data sets, but \name{} is no more impacted by this than Standard Practice(tm)!

There are (at least) three attitudes for an investigator to take in the face of variations in the data, from epoch to epoch or exposure to exposure:
The first is that the procedures we design are designed to deliver the exposure (or signal-to-noise-squared) weighted mean of the observed epochs.
That is, the fact that the data are varying does not mean that the methodology needs to be modified; without modification, the procedures given above---both \name{} and Standard Practice(tm)---will produce something like the mean spectrum.
With this attitude, the spectral variations are just another source of noise.

The second attitude to take---and the necessary attitude for many projects---is that the variations must be accounted for, by complexifying the forward model.
For example, if there is a varying line-spread function between epochs (as there is for the \textsl{APOGEE} spectrograph, because the LSF is a function of the slit position of the optical fiber), then in addition to a shift at each epoch there is also an LSF adjustment at each epoch.
The LSF adjustment is then applied to the spectral model before it is compared to each epoch of data, just as the shift is applied in the current methodological scope.
Similar model adjustments can account for intrinsic spectral variations, additive or multiplicative nuisance signals in the data (like backgrounds or telluric absorptions), and mistakes in data pre-processing (like continuum modeling errors).

Although \emph{conceptually} it is not hard to account for such variabilities, \emph{in practice} they can make life hard.
For one, the method we have presented assumes that we \emph{know} the individual epoch shifts \textsl{a priori}, and all new epoch-to-epoch adjustments would also depend on per-epoch housekeeping data or parameters that might, in practice, be hard to know.
That brings us to the third attitude that the investigator can take.
What's presented above can be seen as a model for the \emph{mean} of the process that generates individual-epoch spectra; that is the first term in an expansion of moments.
The next term would be the \emph{variance} of the process.
That is, a more sophisticated model could have not just $P$ parameters for the mean spectrum, but additionally $P'$ parameters for a variance model.
If the variance is low dimensional, or sparse, or compact in important ways, there should be very effective models of this form.
For example, we could imagine forward-modeling generalizations of some of sparse methods (for example, \citealt{candes}) for applications in which the variability is localized in time or space; or we could imagine generalizations of matrix factorizations (for example, \citealt{hmf}) for applications in which variability is low dimensional (the \textsl{wobble} method has some such functionality; \citealt{wobble}).

The combined spectrum created by \name{} has minimal pixel-to-pixel covariances based on repeated experiments.
In contrast, the combined spectrum created by Standard Practice(tm) has substantial pixel-to-pixel covariances.
This is particularly evident in the second panel of Figure~\ref{fig:standard} (say near $\ln\lambda = 8.7024$).
Here the pixel-to-pixel covariances resulting from Standard Practice(tm) are not purely of academic interest, as the covariances here mimic weak spectral features.
For example, if a weak Na line existed near this wavelength then a spectroscopist might readily interpret these covariances as real spectral features and report an enhanced Na abundance.
Checking the strength of other (stronger) Na features would be a sensible check for this interpretation, but often there are few spectral lines available for a given element.

One side note about epoch-to-epoch variations in instrument or data-processing parameters:
It is usually a good idea to model not the full instrument or data properties (as does \citealt{sp}) if that can be avoided, but rather just the \emph{differences} in instrument or data properties from one epoch to the next, taking those differences away from some fiducial data or instrument state.
In the case of the LSF, for example, that obviates the need to model the infinite-resolution spectrum; the spectrum only needs to be modeled at fiducial resolution and convolved to other resolutions with difference kernels.

All the above considerations of variability bring us to another of our strong assumptions:
We have assumed that we know the shifts $\Delta x_i$ exactly.
In practice, we rarely do, or maybe more accurately: We rarely know them as well as we could know them, in principle.
When the shifts $\Delta x_i$ are not known, or not known accurately enough, we recommend a simultaneous-fitting procedure, in which the shifts are learned at the same time as the combined spectrum is estimated or inferred.
This can be achieved through a challenging nonlinear optimization, or through an iteration of linearized optimizations, alternating improvements to the combined spectrum and improvements to the shifts.
This is, more or less, how the \textsl{wobble} system works (\citealt{wobble}).

We effectively assumed that the noise was Gaussian, and zero mean, but these are not always true of astronomical data.
In principle even in the best-case scenarios there are Poisson noise sources, and cosmic rays and faulty detector pixels tend to add in highly skewed, non-zero-mean noise.
The most important noise assumption made in all this is the zero-mean assumption; when this is wrong, all current spectral (or image) combination methods will be biased (including Standard Practice), unless they employ a likelihood function that accurately represents the detailed noise statistics.

All code used for this project is available at \url{https://github.com/davidwhogg/NoDataInterpolation}.

\paragraph{Software}
\texttt{numpy} \citep{numpy} ---
\texttt{scipy} \citep{scipy} ---
\texttt{matplotlib} \citep{matplotlib}.

\begin{acknowledgments}
It is a pleasure to thank
Mike Blanton (NYU),
Matt Daunt (NYU),
Dylan Green (UC Irvine),
Jon Holtzman (NMSU), 
Dustin Lang (Perimeter),
Adrian Price-Whelan (Flatiron),
and the Astronomical Data Group at the Flatiron Institute
for valuable discussions and input.
The Flatiron Institute is a division of the Simons Foundation.
This research was supported in part by the Australian Research Council Centre of Excellence for All Sky Astrophysics in 3 Dimensions (ASTRO 3D), through project number CE170100013.
\end{acknowledgments}


\begin{thebibliography}{dummy}
\bibitem[Barnett et al.(2019)]{finufft}
Barnett, A. H., Magland, J. F., \& af~Klinteberg, L., 2019, SIAM J. Sci. Comput. 41(5), C479
\bibitem[Bedell et al.(2019)]{wobble} Bedell, M., Hogg, D.~W., Foreman-Mackey, D., et al.\ 2019, \aj, 158, 164. doi:10.3847/1538-3881/ab40a7
\bibitem[Bolton \& Schlegel(2010)]{sp} Bolton, A.~S. \& Schlegel, D.~J.\ 2010, \pasp, 122, 248. doi:10.1086/651008
\bibitem[Candes \& Romberg(2007)]{candes} Candes, E., \& Romberg, J., 2007, Inverse problems, 23, 969.
\bibitem[Fruchter \& Hook(2002)]{drizzle} Fruchter, A.~S. \& Hook, R.~N.\ 2002, \pasp, 114, 144. doi:10.1086/338393
\bibitem[Guy et al.(2023)]{desipipeline} Guy, J., Bailey, S., Kremin, A., et al.\ 2023, \aj, 165, 144. doi:10.3847/1538-3881/acb212
\bibitem[Harris et al.(2020)]{numpy} Harris, C.~R., Millman, K.~J., van der Walt, S.~J., et al.\ 2020, \nat, 585, 357. doi:10.1038/s41586-020-2649-2
\bibitem[Hastie et al.(2022)]{doubledescent} Hastie, T., Montanari, A., Rosset, S., \& Tibshirani, R. J., 2022, Annals of Statistics, 50, 949. doi:10.1214/21-AOS2133
\bibitem[Hewett et al.(1985)]{oe} Hewett, P.~C., Irwin, M.~J., Bunclark, P., et al.\ 1985, \mnras, 213, 971. doi:10.1093/mnras/213.4.
\bibitem[Hogg et al.(2010)]{fitting} Hogg, D.~W., Bovy, J., \& Lang, D.\ 2010, arXiv:1008.4686
\bibitem[Hogg \& Villar(2021)]{fittingflexible} Hogg, D.~W. \& Villar, S.\ 2021, \pasp, 133, 093001. doi:10.1088/1538-3873/ac20ac
\bibitem[Hunter(2007)]{matplotlib} Hunter, J.~D.\ 2007, Computing in Science and Engineering, 9, 90. doi:10.1109/MCSE.2007.55
\bibitem[Ivezi{\'c} et al.(2019)]{lsst} Ivezi{\'c}, {\v{Z}}., Kahn, S.~M., Tyson, J.~A., et al.\ 2019, \apj, 873, 111. doi:10.3847/1538-4357/ab042c
\bibitem[Kelson(2003)]{kelson} Kelson, D.~D.\ 2003, \pasp, 115, 688. doi:10.1086/375502
\bibitem[Lang et al.(2009)]{undetectable} Lang, D., Hogg, D.~W., Jester, S., et al.\ 2009, \aj, 137, 4400. doi:10.1088/0004-6256/137/5/4400
\bibitem[Majewski et al.(2017)]{apogee} Majewski, S.~R., Schiavon, R.~P., Frinchaboy, P.~M., et al.\ 2017, \aj, 154, 94. doi:10.3847/1538-3881/aa784d
\bibitem[Tsalmantza \& Hogg(2012)]{hmf} Tsalmantza, P. \& Hogg, D.~W.\ 2012, \apj, 753, 122. doi:10.1088/0004-637X/753/2/122
\bibitem[Virtanen et al.(2020)]{scipy} Virtanen, P., Gommers, R., Oliphant, T.~E., et al.\ 2020, Nature Methods, 17, 261. doi:10.1038/s41592-019-0686-2
\bibitem[Voinov \& Nikulin(2012)]{mvue} Voinov, V. G., \& Nikulin, M. S., 2012, Unbiased Estimators and Their Applications: Volume 1: Univariate Case, Springer Science \& Business Media.
\bibitem[Williams et al.(1996)]{hdf} Williams, R.~E., Blacker, B., Dickinson, M., et al.\ 1996, \aj, 112, 1335. doi:10.1086/118105
\bibitem[Wilson et al.(2019)]{apogeehardware} Wilson, J.C., Hearty, F.R., Skrutskie, M.F., Majewski, S.R., Holtzman, J.A., Eisenstein, D., et al.\ 2019, \pasp, 131, 055001. doi:10.1088/1538-3873/ab0075.
\bibitem[Wolf et al.(2003)]{combo17} Wolf, C., Meisenheimer, K., Rix, H.-W., et al.\ 2003, \aap, 401, 73. doi:10.1051/0004-6361:20021513
\bibitem[Zackay, B. \& Ofek (2017)]{zackay} Zackay, B. \& Ofek, E. O.\ 2017, \apj, 836, 188. doi:10.3847/1538-4357/836/2/188
\bibitem[Zechmeister et al.(2014)]{froe} Zechmeister, M., Anglada-Escud{\'e}, G., \& Reiners, A.\ 2014, \aap, 561, A59. doi:10.1051/0004-6361/201322746

\end{thebibliography}
\end{document}